%% file: attenuation_v2_6_1_jinst.tex
\documentclass[a4paper,11pt]{article}
\pdfoutput=1 

\usepackage{jinstpub} 

\usepackage{graphicx}
\usepackage{natbib}
\usepackage{array}
\usepackage[%
  colorlinks=true,
  urlcolor=blue,
  linkcolor=blue,
  citecolor=blue
]{hyperref}

\title{\boldmath In Situ Measurement of Relative Attenuation Length of 
Gadolinium-Loaded Liquid Scintillator Using Source Data at RENO Experiment}

\include{RENO_authors_2016_JINST_v3}

\abstract{
We present in situ measurements of the relative attenuation length of the gadolinium-loaded 
liquid scintillator in the RENO (Reactor Experiment Neutrino Oscillation) detectors 
using radioactive source calibration data. 
We observed a steady decrease in the attenuation length of the Gd-LS in the RENO detectors 
by $\sim$50\% in about four years since the commissioning of the detectors.
}


\arxivnumber{1609.09483}

\collaboration{RENO collaboration}

\begin{document} 
\maketitle
\flushbottom

\section{Introduction}\label{sec:intro}
Organic solvent-based liquid scintillators (LS) have been commonly employed 
for low-energy neutrino experiments due to their effectiveness. However,
these experiments typically last from several months to years, the long-term 
stability of the LS is of utmost importance.
Any changes in the light yield or attenuation length of LS requires extensive
monitoring and corrections, making the experiment significantly more challenging. 
Unfortunately, there have been reports of the degradation of the attenuation 
length of the gadolinium-loaded LS (Gd-LS) in a relatively short time~\cite{Apollonio:1997xe, 
Apollonio:2002gd,Almazan:2022aa,JYKim:2022aa}.
The RENO experiment also has experienced degradation of the attenuation length of 
its Gd-LS~\cite{Seo:2016eom}. 

The attenuation length of LS in a detector can be measured by extracting LS samples from the detector
and analyzing them with spectrophotometers or purposefully-built instruments. Commercially
available spectrophotometers typically use light sources with various wavelengths and measure 
the attenuation length of the liquid samples in small cuvettes. However, due to the 
limited sample size, typically a few centimeters, the uncertainty of the attenuation length 
measurement can be significant if the attenuation length exceeds several meters~\cite{Park:2013nsa}.
As a result, spectrophotometers may not be suitable for measuring the attenuation length
accurately in some cases.
On the other hand, specifically built instruments to measure the attenuation length, such 
as the one used in~\cite{Gao:2013pua}, have achieved a sub-meter uncertainty measurements for 
LS with an order of a 10~m attenuation length.
However, these instruments require several liters of LS to make a measurement, limiting the 
frequency of the measurements on fresh LS samples extracted from the detector.
Moreover, once the LS sample is extracted from the detector, the environmental conditions, 
such as temperature, humidity, or oxygen level, may differ from those in the detector.
As a result, the attenuation length measured for the LS sample kept outside the detector
for an extended period could differ from that of the LS in the detector unless
the environmental conditions are closely matched. 

Alternatively, the attenuation length of LS can be measured in situ using radioactive
sources placed in the detector~\cite{Apollonio:2002gd}. 
In this case, the measured quantity is an effective attenuation length as the 
optical photons with different wavelengths are summed over.
The measured attenuation length is for the same optical photon spectrum as the scintillation 
signal in the experiment.
In principle, this method requires a detailed photomultiplier tube (PMT) performance model for each PMT used for the 
measurement. Using the model can be avoided by taking ratios of PMT hits spatially and temporally.
In this paper, we report on the method of measurements of the relative effective attenuation length of 
Gd-LS in the RENO detectors using an in situ method, and the results of the measurements using 
calibration radioactive source data samples spanning about 1\,400 days. 

\section{RENO Detector and Calibration System}
\begin{figure}[htbp]
\centering
\includegraphics[width=9cm]{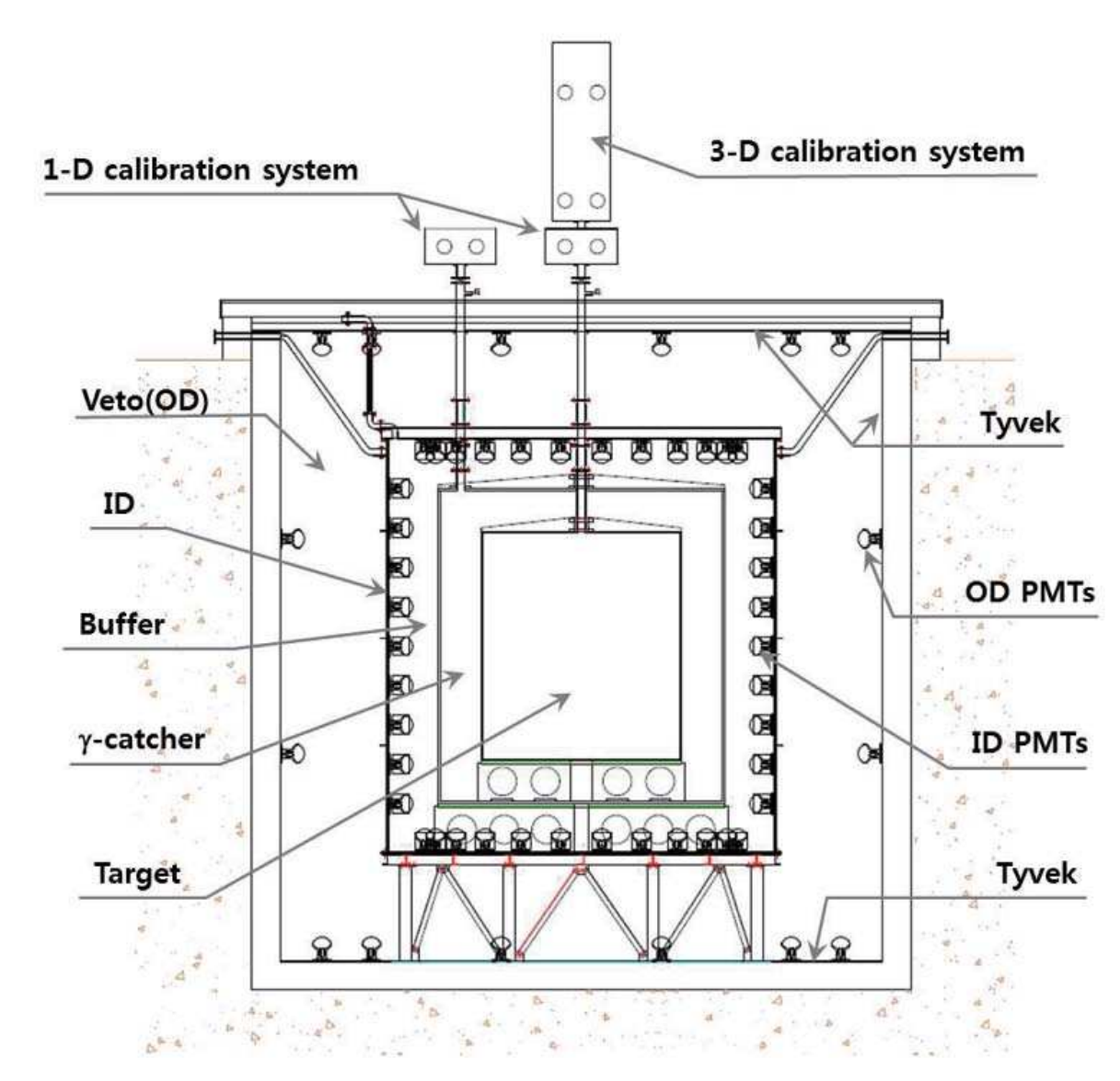}
\caption{Schematic view of the RENO detector. There are two 1-dimensional calibration 
systems each for target and $\gamma$-catcher, that deploy radioactive sources 
vertically ($z$-direction).}
\label{reno detector}
\end{figure}

The RENO is an neutrino oscillation 
experiment located at Hanbit nuclear power plant in Korea.
The RENO detector is described in details in~\cite{Ahn:2010vy} and 
only the relevant part will be briefly described here.
The experiment conists of two identical detectors, near and far detectors.
Figure~\ref{reno detector} shows a schematic view of RENO detector.
Each detector consists of inner and outer detectors, with the outer detector serving
as a cosmic muon veto system. The inner detector is contained in a cylindrical
stainless steel vessel with two nested concentric transparent acrylic 
vessels. The innermost vessel holds 16~tons of linear alkylbenzene (LAB) based 
Gd-LS~\cite{Park:2013nsa} as a target, with a Gd concentration 0.1\%. 
A 60~cm-thick layer of unloaded LS, called the $\gamma$-catcher, is placed between
the target vessel and the outer vessel to recover energy leakage from the target.
The thicknesses of the target and $\gamma$-catcher vessel walls are 2.5~cm
and 3.0~cm, respectively.
A 70~cm-thick buffer region filled with 65 tons of mineral oil (MO) surrounds the 
$\gamma$-catcher. The light signals from the inner detector are detected 
by 354 Hamamatsu R7081 10-inch PMTs mounted perpendicularly on 
the inner wall of the stainless steel vessel and immersed in the buffer. 
The gain of each PMT is measured before the installation and continually
corrected for the gain drift. 
The refractive indices for Gd-LS, LS, MO, and acrylic vessels
at 405~nm are measured to be 1.50, 1.50, 1.47, and 1.50, respectively~\cite{Yeo:2010zz,Park:2012dv}. 
Therefore, no significant reflection or refraction is expected at the medium interfaces.
Table~\ref{dimension table} shows the dimensions of the detector.

\begin{table*}[h]
\centering
\caption{Dimensions of detector layers. The layers are in a concentric
cylindrical shape.}
\begin{tabular}{|c|ccc|}\hline
Layer   &Outer Diameter &Outer Height &Material \\
        &(cm)         &(cm)         &         \\\hline
Target  &275.0  &315.0  &Gd-LS \\
Target Vessel   &280.0  &320.0  &Acrylic Plastic \\
$\gamma$-Catcher        &394.0  &434.0  &LS \\
$\gamma$-Catcher Vessel &400.0  &440.0  &Acrylic Plastic \\
Buffer &538.8   &578.8  &MO \\
Buffer Vessel &540.0    &580.0  &Stainless Steel \\
Veto &840.0     &889.8  &Water \\\hline
\end{tabular}
\label{dimension table}
\end{table*}

There are two 1-dimensional source calibration systems available on each detector, 
one for the target and the other for the $\gamma$-catcher. These systems allow for 
deploying a radioactive source in the liquids in the vertical direction ($z$-direction).  
Only the target calibration system is used for this measurement, where the source moves 
along the axial centerline of the target. The system consists of a 
a Teflon-PFA container housing a radioactive source, which is attached at the
end of a wire made of aromatic polyamide that is driven by a stepper motor.
A weight is attached to the source container to counteract the buoyancy
when the source is submerged in the liquid.
The relative $z$-position accuracy of the source is measured to be within 
a few millimeters and the systematic uncertainty 
of a few centimeters is expected due to possible shifts in the $z$-position.

\section{Data and Event Selection}
In this study, the data taken between Nov. 2011 and April 2015 with a $^{60}$Co radioactive 
source are used.   
The source is positioned between $(x,y,z)=(0,0,-120)$ and $(0,0,120)$~cm in the target. Here, $(0,0,0)$ 
represents the center of the target. 
Table~\ref{tab:data list} shows the dates of the data samples taken for this measurement.
\begin{table}
\centering
\caption{List of $^{60}$Co source calibration data used in the measurement. The source is 
positioned between $z=\pm120$~cm. The numbers within the parentheses are the elapsed time 
in days since 2011-08-01, corresponding to the commissioning of the RENO detectors.
}
\begin{tabular}{|c|cc|}\hline
Detector              &Date  &Source $z$-position \\
                      &      &Intervals (cm)       \\\hline
Far                   &2011-11-08 (\space\space\space99)&40\\
                      &2012-07-13 (\, 347)&40\\
                      &2012-10-27 (\, 453) &40\\
                      &2013-12-09 (\, 861) &40\\
                      &2014-01-17 (\, 900) &10\\
                      &2015-04-14 (1\,352) &10\\\hline
Near                  &2011-11-10 (\, 101) &40\\
                      &2012-04-19 (\, 262) &40\\
                      &2012-07-12 (\, 346) &40\\
                      &2012-10-27 (\, 453) &40\\
                      &2013-12-09 (\, 861) &40\\
                      &2014-01-18 (\, 901) &10\\
                      &2015-04-13 (1\,351) &10\\\hline
\end{tabular}
\label{tab:data list}
\end{table}

\begin{figure*}[h]
\begin{center}
\includegraphics[width=6cm]{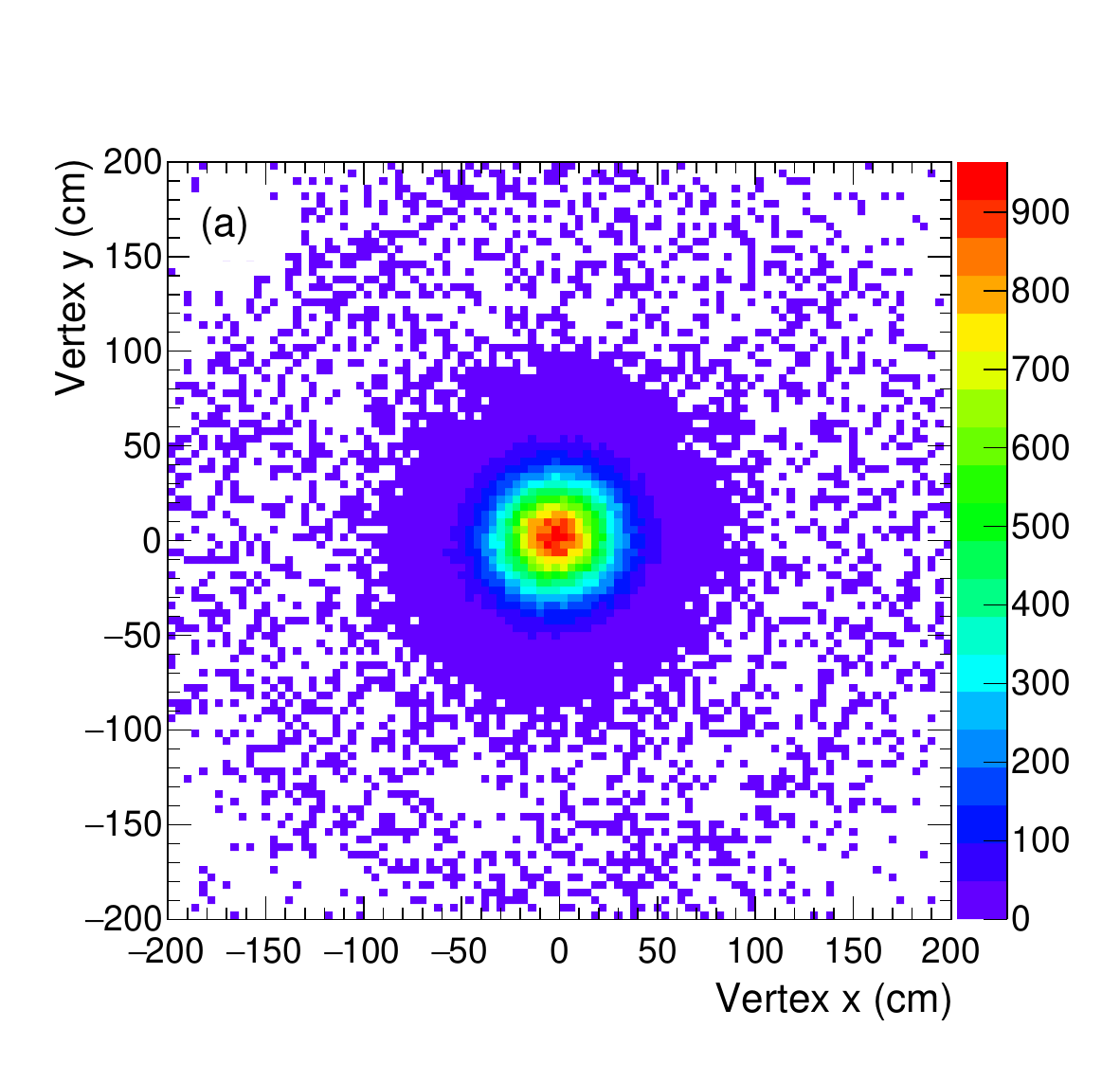}
\includegraphics[width=6cm]{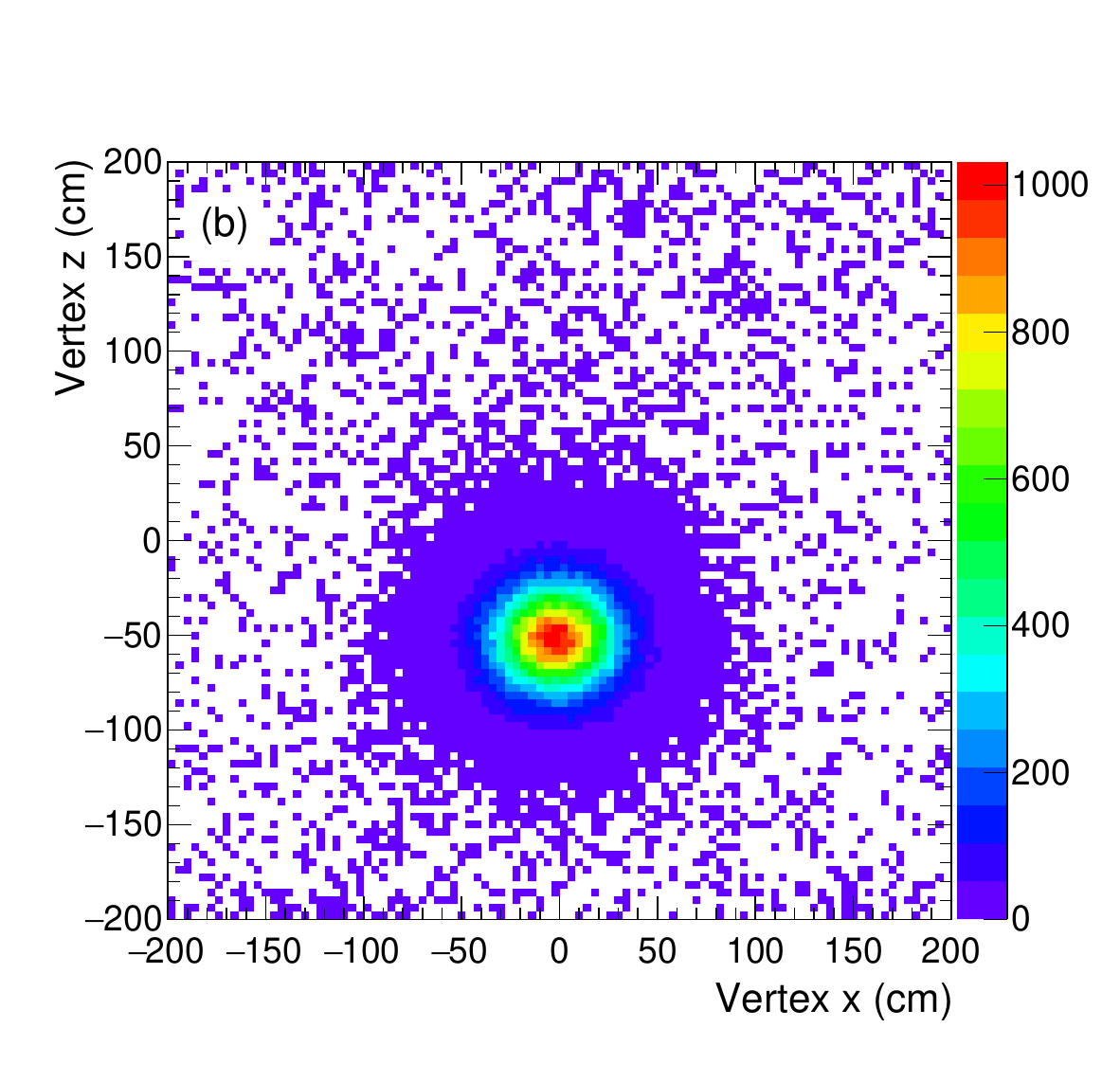}
\includegraphics[width=6cm]{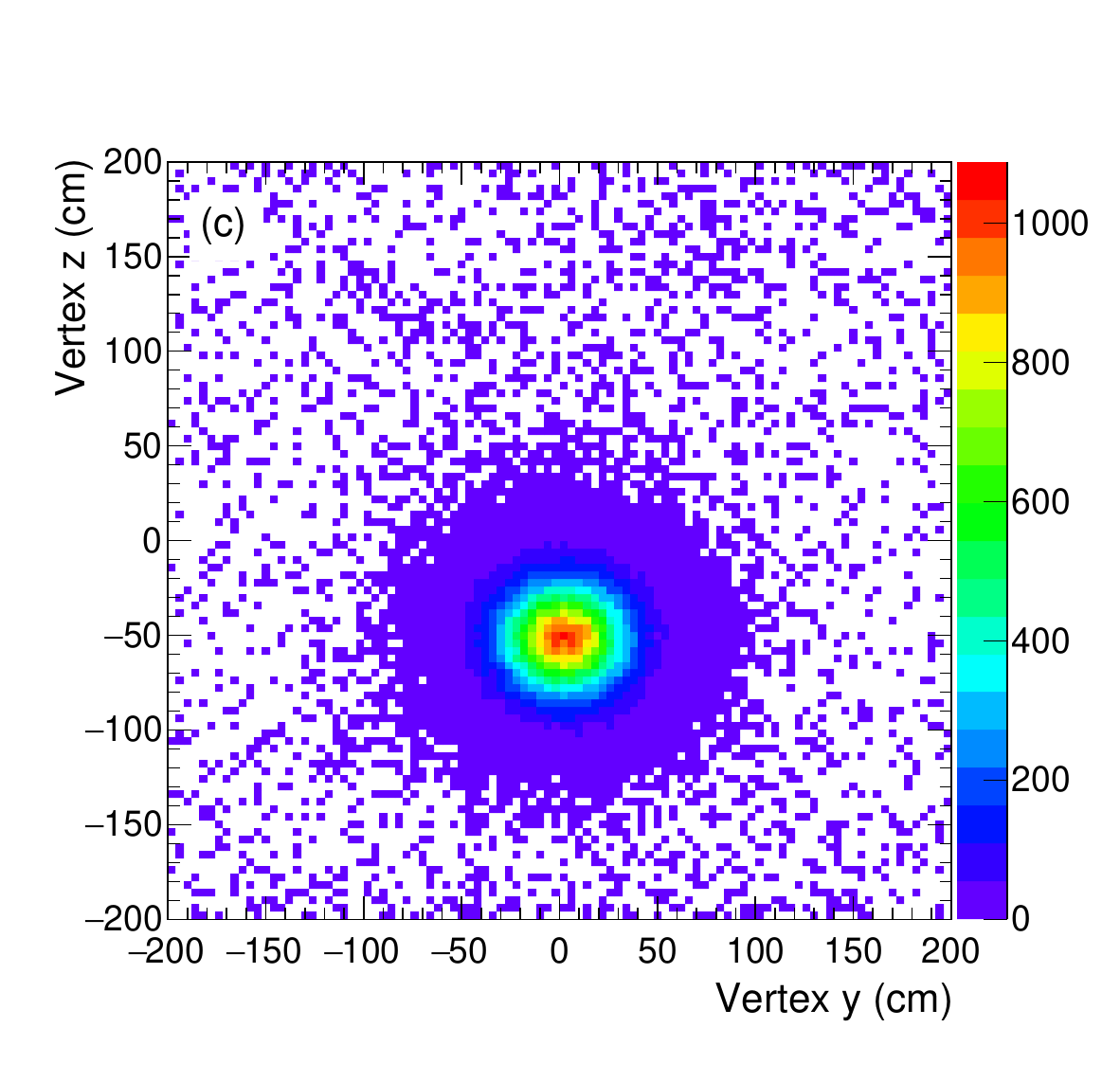}
\end{center}
\caption{The reconstructed event vertex distributions in (a) $x$-$y$, (b) $x$-$z$, and (c) $y$-$z$ 
of the data taken at 2014-01-17 at far detector. A $^{60}$Co source is located at a nominal $z$-position 
of $-50$~cm. The widths of the distributions are approximately $20$~cm in all directions.}
\label{fig:vertex 2d dist}
\end{figure*}

The event vertex was reconstructed for each event using the weighting method
described in~\cite{Kim:2012cuq}. Figure~\ref{fig:vertex 2d dist} shows the
reconstructed event vertex distributions of a calibration data sample for the far detector.
The event vertex distributions in $x$-$y$, $y$-$z$, and $z$-$x$ are fit with 2-dimensional 
Gaussian functions simultaneously. 
The event selection criterion used is based on the reconstructed vertex position of each
event, denoted by ($x$, $y$, $z$). Only events that satisfy the following condition are used 
for the analysis:
\begin{equation}
         \frac{(x-\bar{x})^2}{\sigma_x^2}
        +\frac{(y-\bar{y})^2}{\sigma_y^2}
        +\frac{(z-\bar{z})^2}{\sigma_z^2} <1,
\label{eq:vertex requirement}
\end{equation}
where $(\bar{x},\bar{y},\bar{z})$ and ($\sigma_x$, $\sigma_y$, $\sigma_z$) are the
means and widths, respectively, of the fits.
This event vertex criterion ensures that the event occurred at the desired vertex position 
and removes most of background events. No other event selection requirement is used.

\section{Method of Relative Attenuation Length Measurement}
When a radioactive source is submerged in a LS, particles from the source 
lose their energy in the immediate vicinity of the source and optical photons from the 
scintillating process are radiated isotropically.
If a light source emits $N$ optical photons isotropically in a uniform medium at $\vec{r}$ from a PMT
pointing in $\hat{\xi}$ direction, then the number of photons detected by the PMT, $n$, can be written as
\begin{equation}
n = N\,\frac{\Omega(\vec{r},\hat{\xi})}{4\pi}\,\epsilon(\vec{r},\hat{\xi})\,\exp(-r/\lambda),
\end{equation}
where $\Omega(\vec{r},\hat{\xi})$ is the PMT coverage in solid angle, $\epsilon(\vec{r},\hat{\xi})$ is the 
PMT efficiency, and $\lambda$ is the attenuation length of the medium.
Here, $\lambda$ is the effective attenuation length for scintillating photons 
but not of the specific wavelength photons. It takes into account any effects 
experienced by optical photons in a medium, including 
reflection of the photons at the boundary of the medium.

\begin{figure}
\begin{center}
\includegraphics[width=10cm]{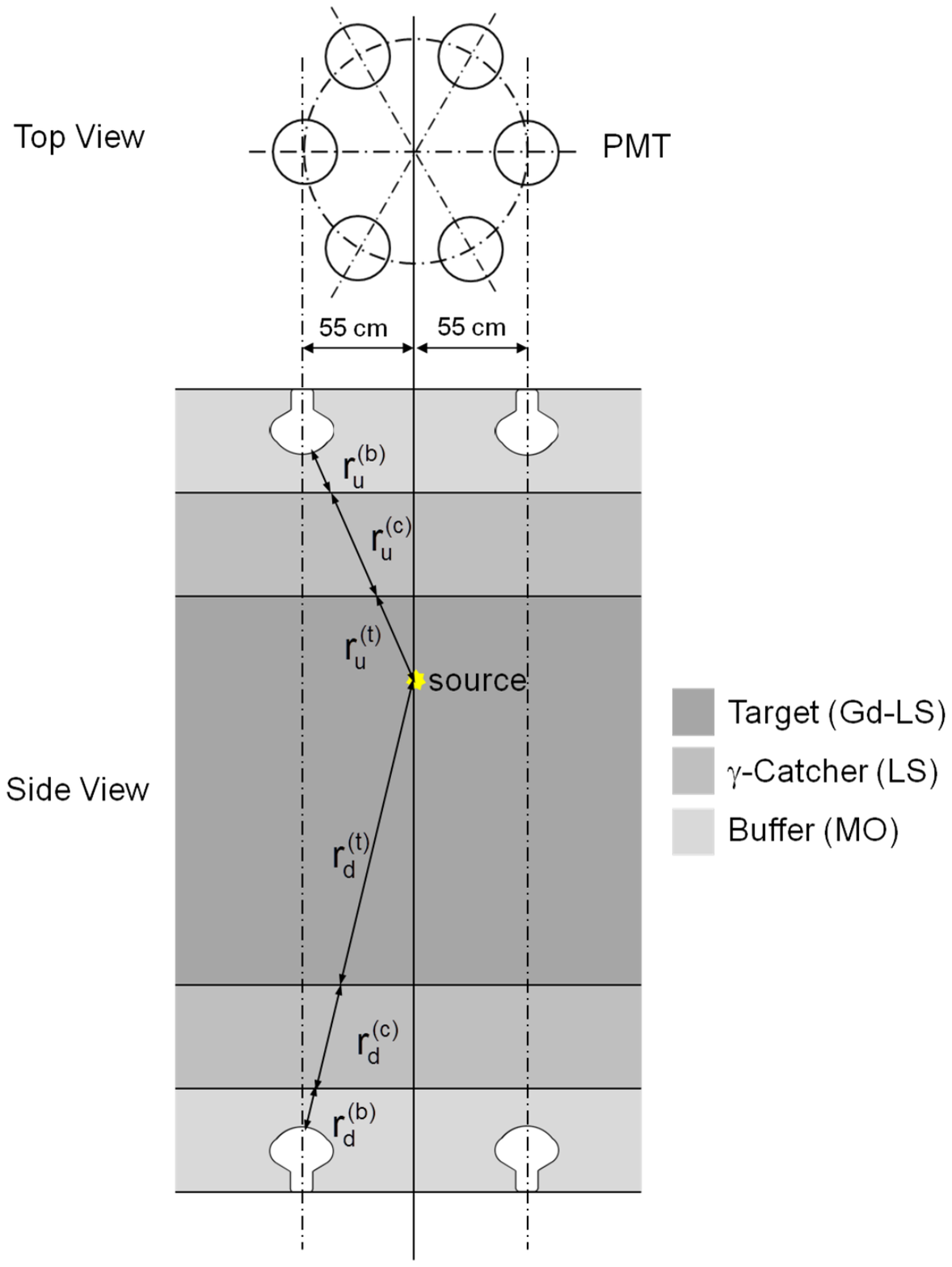}
\end{center}
\caption{The location of six PMTs (solid line circle) nearest to the
centerline of the detector at the top ($u$) and bottom ($d$) ends of the inner
detector from above (not drawn to scale). They are equally spaced at 55~cm from the
centerline where calibration sources are deployed. The distance
from any of the PMTs on either top or bottom and to the calibration
source is the same. The path lengths between the radioactive source and the
top (bottom) PMT in target, $\gamma$-catcher, and buffer are $r_{u(d)}^{(t)}$, 
$r_{u(d)}^{(c)}$, and $r_{u(d)}^{(b)}$, respectively.}
\label{fig:PMT arrangement}
\end{figure}

The optical photons generated in the target must traverse three layers of liquids in the target (``$t$''),
$\gamma$-catcher (``$c$''), and buffer (``$b$'') as well as two layers of acrylic vessel walls to reach
any PMT in the inner detector. 
The ratio of numbers of photons detected by any two PMTs $i$ and $j$ is 
\begin{eqnarray}
\frac{n_i}{n_j}&=&\frac{\Omega_i\,\epsilon_i \prod_{\ell}\exp(-r_i^{(\ell)}/\lambda^{(\ell)}) }
                       {\Omega_j\,\epsilon_j \prod_{\ell}\exp(-r_j^{(\ell)}/\lambda^{(\ell)}) } 
               \nonumber \\
               &=&\frac{\Omega_i\,\epsilon_i}{\Omega_j\,\epsilon_j}
                  \exp\left(-\sum_{\ell}\frac{\Delta r_{ij}^{(\ell)}}{\lambda^{(\ell)}}\right), 
\label{eq:fraction}
\end{eqnarray}
where $\Delta r_{ij}^{(\ell)}\equiv r_i^{(\ell)}-r_j^{(\ell)}$ in layer $\ell$,
$\Omega_i\equiv\Omega(\vec{r}_i,\hat{\xi}_i)$, and $\epsilon_i\equiv\epsilon(\vec{r}_i,\hat{\xi}_i)$.
Here $\hat{\xi}_i$ is the unit vector in the direction that the $i^{\rm th}$ PMT is pointing.
If $|\Delta r^{(\ell)}/\lambda^{(\ell)}|$ ($\ell\neq t$) can be made much
smaller than $|\Delta r^{(t)}/\lambda^{(t)}|$ then eq.~(\ref{eq:fraction}) can be approximated
in terms of the target properties as
\begin{equation}
\frac{n_i}{n_j} \simeq \frac{\Omega_i \epsilon_i}{\Omega_j \epsilon_j}
                 \exp\left(-\frac{\Delta r_{ij}^{(t)}}{\lambda^{(t)}}\right). 
\label{eq:approx eqn}
\end{equation}
If $({\Omega_i \epsilon_i}/{\Omega_j \epsilon_j})$ does not change with respect to $\Delta r$,
one can measure $\lambda^{(t)}$ by measuring $n_i$, $n_j$, and $\Delta r_{ij}$. 
A similar technique has been used to measure the attenuation length of Gd-LS in 
at CHOOZ experiment~\cite{Apollonio:2002gd} where the CHOOZ detector has only the target to consider. 
For the RENO detectors this condition can be satisfied using two sets of six PMTs each installed at top and bottom
closest to the axial centerline of the target, where the target calibration sources are deployed
along. The schematic of the arrangement of these PMTs relevant to this study is shown
in figure~\ref{fig:PMT arrangement}. For the source positions
at $z=\pm 120$~cm, where the effects from the other layers are the greatest in the data used in
this study, the deviations in $n_i/n_j$ coming from the other layers are estimated to be less than 1\%.
The effects of the uniform changes in attenuation lengths 
in non-target layers over time to the measurement should be negligible as well.
Since each of the six PMTs at either top or bottom used for the measurements has the same 
distance and polar angle with respect to the radioactive source, they are treated as a single PMT. 
Henceforth, the top six PMTs shall be labeled as ``$u$'' and the bottom ones as ``$d$.''
Although using multiple PMTs makes the measurement by eq.~(\ref{eq:approx eqn})
less sensitive to the characteristics of individual PMT by averaging them out,
however, the measurement is still found to be sensitive to the $\epsilon(\vec{r},\hat{\xi})$ 
model.

Introducing the time dependence to eq.~(\ref{eq:approx eqn}) and defining the ratio 
as $R(t,\Delta r)$ give
\begin{eqnarray}
R(t,\Delta r)&\equiv&\frac{n_u(t)}{n_d(t)} \nonumber\\
         &=&\frac{\Omega_u \epsilon_u(t)}{\Omega_d \epsilon_d(t)}
                 \exp\left(-\frac{\Delta r}{\lambda(t)}\right), 
\label{eq:approx eqn2}
\end{eqnarray}
where $\epsilon_{u(d)}(t) = \epsilon(\vec{r}_{u(d)},\hat{\xi}_{u(d)},t)$
and $\Delta r \equiv \Delta r_{ud}$.\protect\footnote{The superscript 
``${(t)}$'' signifying the parameter to be that of the 
target shall be omitted from henceforth for convenience.} 
The term $\Omega$ is a geometrical term and does not have the time dependence.
Taking the ratio of $R(t,\Delta r)$'s obtained at two different times $t_1$ and $t_2$ gives
\begin{equation}
\frac{R(t_1,\Delta r)}{R(t_2,\Delta r)}=
\frac{\epsilon_u(t_1)\epsilon_d(t_2)}{\epsilon_u(t_2)\epsilon_d(t_1)}
   \exp\left(-\frac{\Delta r}{\lambda_m(t_1,t_2)}\right),
\label{eq:approx eqn3}
\end{equation}
where
\begin{equation}
\frac{1}{\lambda_{m}(t_1,t_2)}\equiv\frac{1}{\lambda(t_1)}-\frac{1}{\lambda(t_2)}.
\label{eq:relative attenuation}
\end{equation}
Here, $\Omega_{u(d)}$ is a geometrical factor and does not change over time. Therefore, $\Omega_{u(d)}$ 
terms cancel out in the equation. If the angular efficiency profile of a PMT remains the same over time, then 
$\epsilon_{u(d)}(t_1)/\epsilon_{u(d)}(t_2)$ at any radioactive source position becomes constant 
for any time interval between $t_1$ and $t_2$.
The coefficient of the exponential of $-\Delta r/\lambda_m(t_1,t_2)$, which 
contains the PMT performance model, becomes a constant coefficient with respect to 
the source position, i.e., with respect to $\Delta r$.
Therefore, if either $\lambda(t_1)$ or $\lambda(t_2)$ is known, the other can be calculated
from the measured $\lambda_m(t_1,t_2)$. 

\begin{figure*}[htbp!]
\centering
\includegraphics[width=7cm]{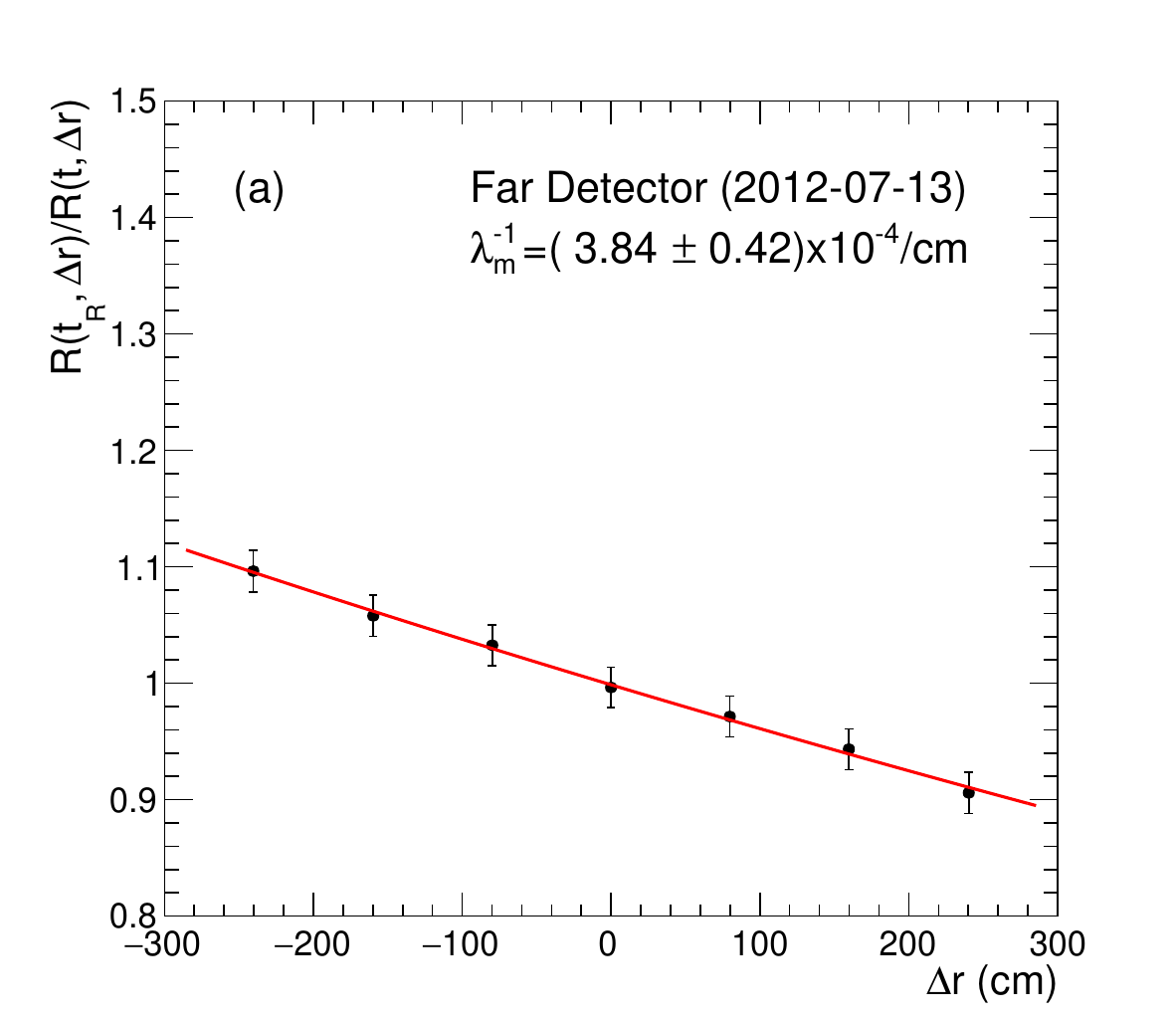}
\includegraphics[width=7cm]{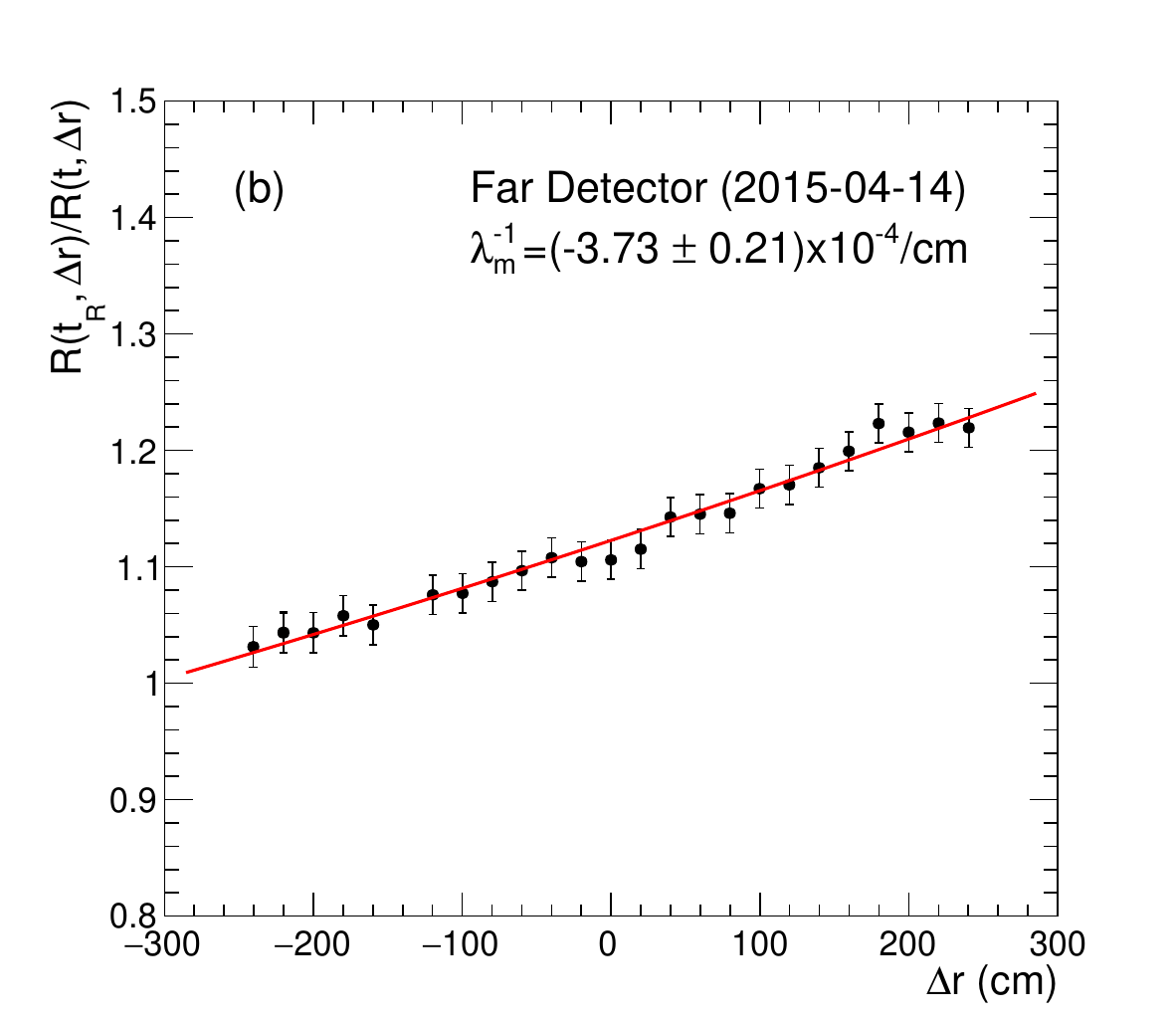}
\\
\includegraphics[width=7cm]{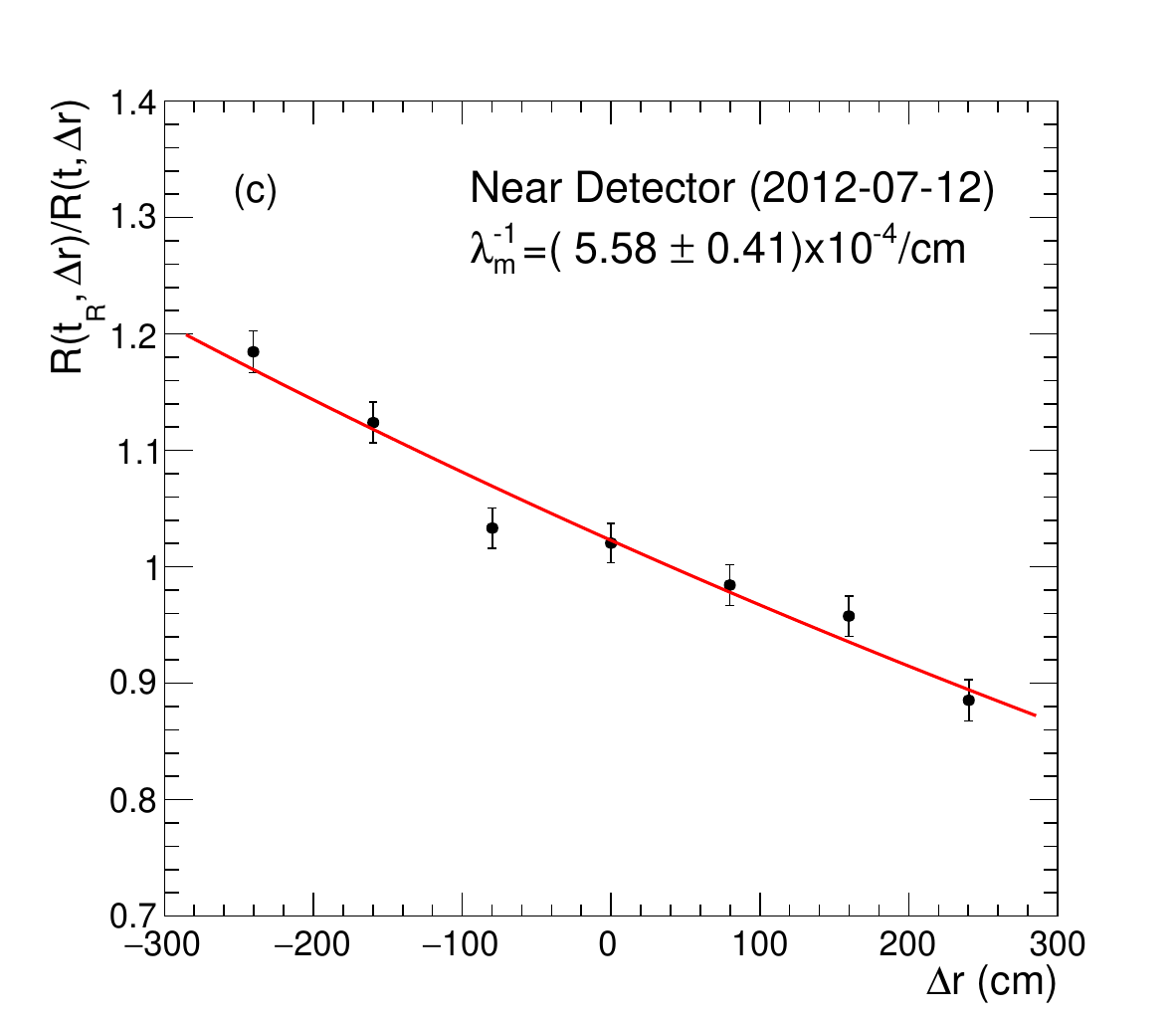}
\includegraphics[width=7cm]{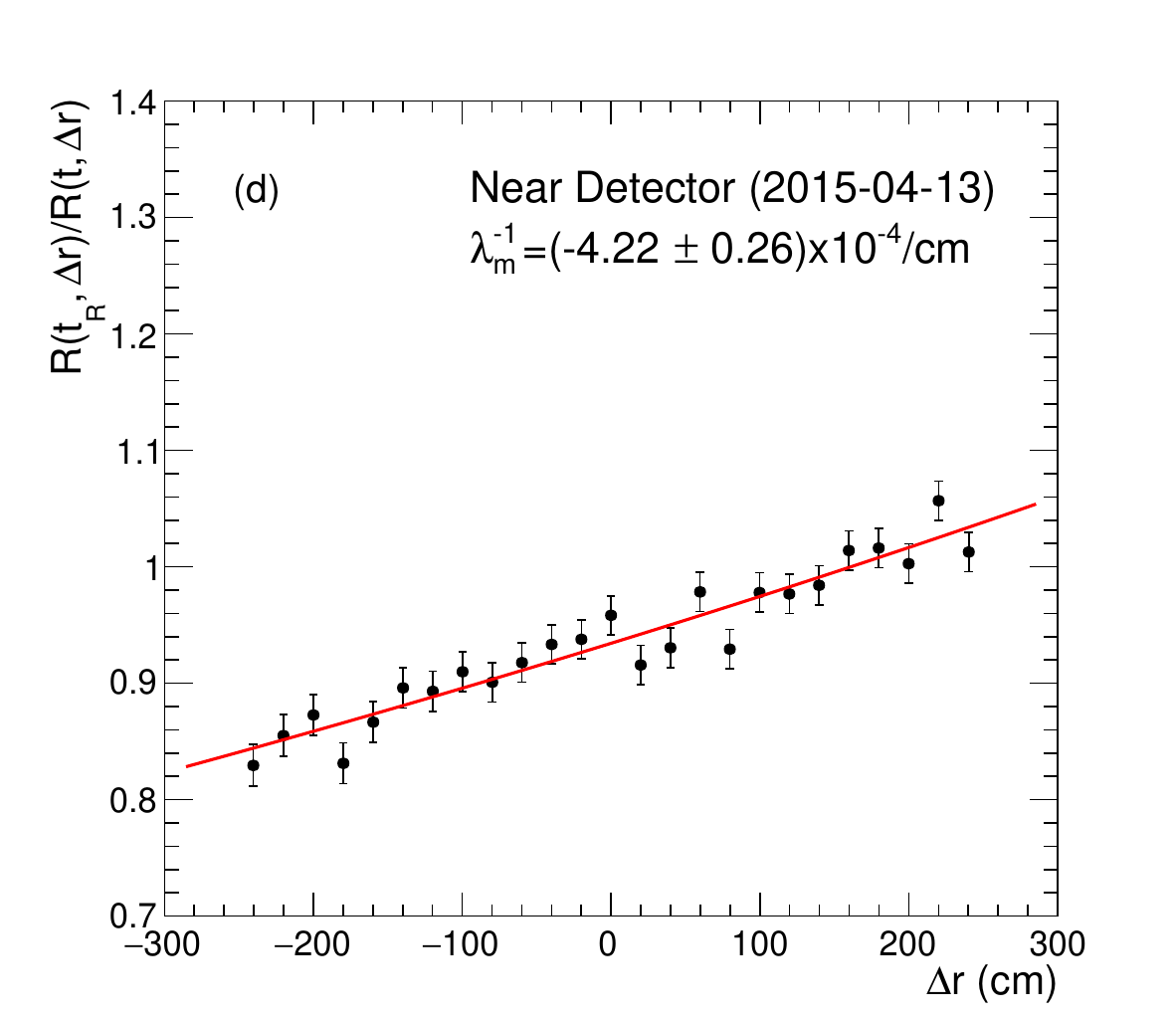}
\caption{The ratio of ratios $R(t_R,\Delta r)/R(t,\Delta r)$ (see eq.~(\protect\ref{eq:approx eqn3})) for
(a) far detector at 2012-07-13 ($t=347$ days), 
(b) at 2015-04-12 (1351 days),
(c) near detector at 2012-07-12 (346 days), 
and (d) at 2015-04-13 (1351 days) and $t_R=900~(901)$~days at 2014-01-17 (2014-01-18) since 2011-08-01
for far (near) detector, fit with an exponential function.}
\label{fig:number ratio3}
\end{figure*}

\begin{figure}[htbp!]
\centering
\includegraphics[width=8cm]{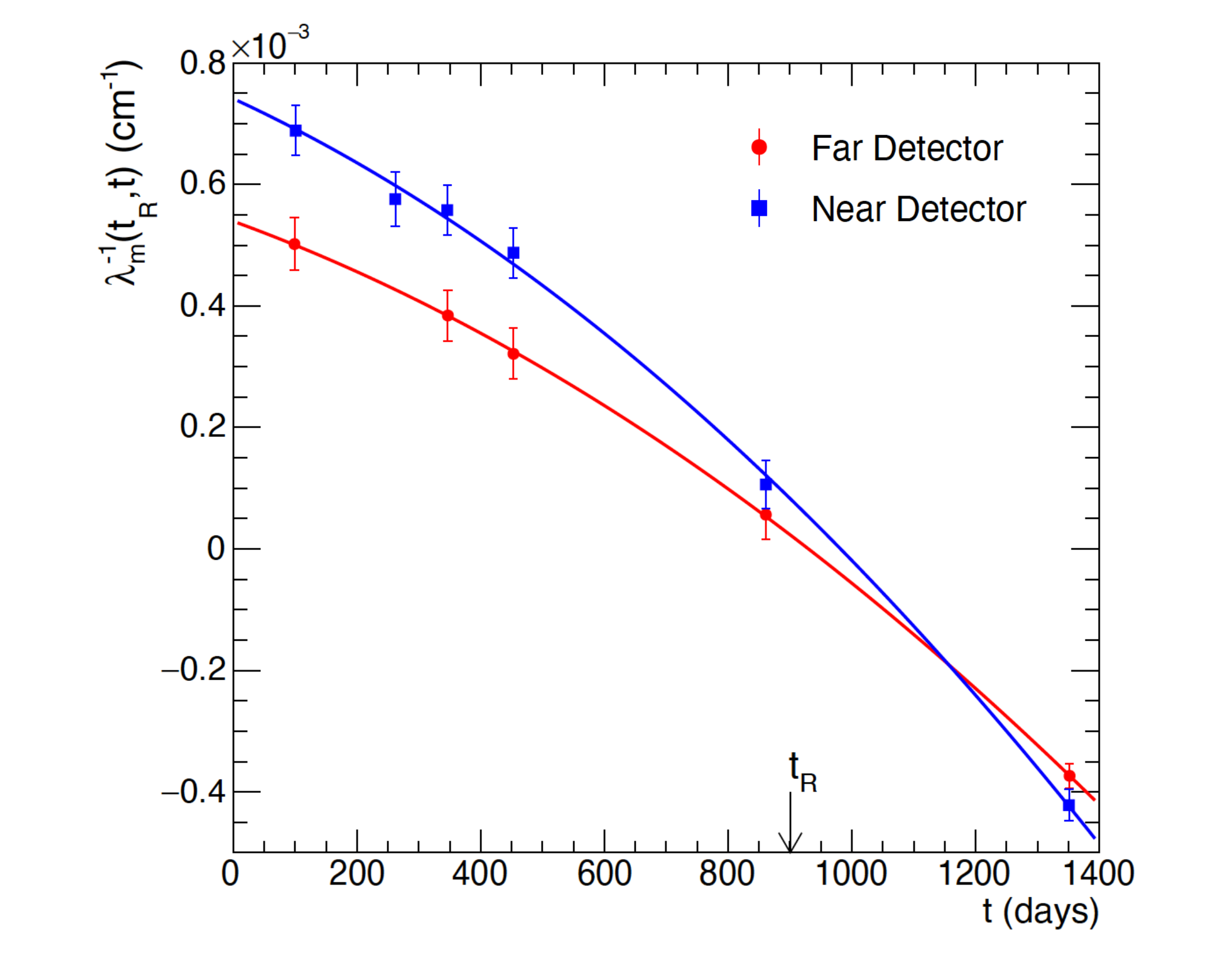}
\caption{The relative inverse of attenuation lengths $1/\lambda_m(t_R,t)$
(see eq.~(\protect\ref{eq:relative attenuation})) as a function of the days since 2011-08-01 for
far (red circles) and near (blue squares) detectors. The slope of $R(t_R,\Delta r)/R(t,\Delta r)$ 
represents the difference between
the inverses of the attenuation length at a given time $t$ in days and at the reference time $t_R$ 
at 2014-01-17 (2014-01-18) for far (near) detector. The solid lines show the second order polynomial fits.
The fitting results are shown in table~\protect\ref{tab:fit results}.}
\label{fig:attenuation results}
\end{figure}

To measure $1/\lambda_m(t_1,t_2)$, we plot $R(t_1,\Delta r)/R(t_2,\Delta r)$ as a function of $\Delta r$ 
using eq.~(\ref{eq:approx eqn3}) with respect to a reference time $t_R = t_1$ against $t_2$.
The time $t$ is calculated as the elapsed time in days since 2011-08-01, corresponding to 
the commissioning date of the detectors.
Here, $t_R$ is chosen to be 900 (901) days for far (near) detector, corresponding to
the fifth (sixth) set of data samples is taken (See table~\ref{tab:data list}).
Figure~\ref{fig:number ratio3} shows $R(t_R,\Delta r)/R(t,\Delta r)$ as a function of 
$\Delta r$, where $t$ is $347$ and $1\,352$ days 
(346 and 1\,351 days) for far (near) detector.\protect\footnote{
Note that a data point is missing near $\Delta r = -180~\rm cm$ in figure~\ref{fig:number ratio3} (b). This is due to a 
missing data point near $\Delta r = -180~\rm cm$ at $t_R$.}
The resulting $1/\lambda_{m}(t_R,t)$ vs $t$ for all data is shown in figure~\ref{fig:attenuation results}
and the fitting results with a second order polynomial is shown in table~\ref{tab:fit results}.
The point at $t=900$ (901) days for the far (near) detector is not shown in figure~\ref{fig:attenuation results} 
and not used in fitting $\lambda_m(t_R,t)$ with a second order polynomial since $\lambda_m(t,t)=0$ by definition.

If the shape of $1/\lambda(t_R,t)$ curve is known, the relative attenuation length can be 
obtained with respect to $t_R$ at any time within the time range of calibration data are taken. 
For example, if the attenuation length Gd-LS of $12.0\pm 0.5$~m at day 0 is assumed for both
far and near detectors, the attenuation lengths are calculated to have been degraded to 
$5.7\pm 0.3$~m and $5.0\pm 0.3$~m for far and near detectors, respectively, over a period of 
about 1\,350 days.

The Gd-LS in the target had been exposed to ambient air for $\sim 1\,400$ days. Since the detectors 
were purged with nitrogen gas and sealed, the degradation of the attenuation length appears to have 
stopped for both the near and far detectors. The cause of this degradation is under investigation,
but it could be attributed to oxidation or humidity.

\section{Summary}
An in situ method is presented for measuring the relative attenuation length for the target Gd-LS 
at RENO experiment using a radioactive source without assuming a PMT efficiency model. It is found 
that the attenuation length of the Gd-LS in the RENO detectors has been degraded by about 50\% over 
1\,350 days since the filling of the detectors, assuming the initial attenuation length of Gd-LS 
at beginning of the experiment is greater than $\sim 10$~m.  

\acknowledgments
The RENO experiment is supported by the National Research Foundation of Korea (NRF) grant No. 
2009-0083526 funded by the Korea Ministry of Science, ICT \& Future Planning. Some of us have 
been supported by a fund from the NRF grant No. 2022R1A5A1030700. We gratefully acknowledge the cooperation of the 
Hanbit Nuclear Power Site and the Korea Hydro \& Nuclear Power Co., Ltd. (KHNP).

\begin{table*}[htbp!]
\centering
\caption{Fitting results of figure~\protect\ref{fig:attenuation results}
with a second order polynomial $\lambda_m^{-1}(t_R,t) = p_0+p_1\,t+p_2\,t^2$,
where $t$ is in days since 2011-08-01.}
\begin{tabular}{|c|ccc|}\hline
Detector &$p_0$ &$p_1$  &$p_2$\\
        &$(\times 10^{-4}~{\rm cm}^{-1})$  &$(\times 10^{-7}~{\rm cm}^{-1}{\rm day}^{-1})$ &$(\times 10^{-10}~{\rm cm}^{-1}{\rm day}^{-2})$\\\hline
Far     &$5.39\pm 0.53$ &$-3.71\pm 1.76$  &$-2.25\pm 1.08$\\
Near    &$7.42\pm 0.48$ &$-4.70\pm 1.70$  &$-2.90\pm 1.07$\\\hline
\end{tabular}
\label{tab:fit results}
\end{table*}
\bibliographystyle{attenuation_v2_6_1_jinst}
\bibliography{attenuation_v2_6_1_jinst}
\end{document}

%% file: RENO_authors_2016_JINST_v3.tex
\author[f,1]{H.~S.~Kim\note{Corresponding author \emailAdd{hyunsookim@sejong.ac.kr}}}
\author[a]{S.~Y.~Kim}
\author[b]{J.~H.~Choi}
\author[g]{W.Q.~Choi}
\author[i]{Y.~Choi}
\author[h]{H.~I.~Jang}
\author[c]{J.~S.~Jang}
\author[a]{K.~K.~Joo}
\author[a]{B.~R.~Kim}
\author[a]{J.~Y.~Kim}
\author[i]{S.~B.~Kim}
\author[e]{W.~Kim}
\author[g]{E.~Kwon}
\author[g]{D.~H.~Lee}
\author[a]{I.~T.~Lim}
\author[b]{M.~Y.~Pac}
\author[d]{I.~G.~Park}
\author[e]{J.~S.~Park}
\author[a]{R.~G.~Park}
\author[g]{H.~Seo}
\author[g,2]{{S.~H.~Seo}\note{Present Address: Center for Underground Physics, Institute of Basic Sciences (IBS), Daejeon 34126, Korea}}
\author[e]{Y.~G.~Seon}
\author[a,3]{{C.~D.~Shin}
\note{Present Address: High Energy Accelerator Research Organization (KEK), Tsukuba, Ibaraki 305-0801, Japan}}
\author[i]{J.~H.~Yang}
\author[a,4]{{I.~S.~Yeo}\note{Present Address: Korea Institute of Science and Technology Information, Daejeon 34141, Korea}}
\author[i]{I.~Yu}
\affiliation[a]{Institute for Universe and Elementary Particles, Chonnam National University, Gwangju 61186, Korea}
\affiliation[b]{Department of Radiology, Dongshin University, Naju 58245, Korea}
\affiliation[c]{GIST College, Gwangju Institute of Science and Technology, Gwangju 61005, Korea}
\affiliation[d]{Department of Physics, Gyeongsang National University, Jinju 52828, Korea}
\affiliation[e]{Department of Physics, Kyungpook National University, Daegu 41566, Korea}
\affiliation[f]{Department of Physics and Astronomy, Sejong University, Seoul 05006, Korea}
\affiliation[g]{Department of Physics and Astronomy, Seoul National University, Seoul 08826, Korea}
\affiliation[h]{Department of Fire Safety, Seoyeong University, Gwangju 61268, Korea}
\affiliation[i]{Department of Physics, Sungkyunkwan University, Suwon 16419, Korea}
\emailAdd{hyunsookim@sejong.ac.kr}